\documentclass[useAMS,usenatbib,referee]{mn2e}

\usepackage{graphicx}

\title[ Black-Hole Accretion Discs and Jets at Super-Eddington Luminosity]{
 Black-Hole Accretion Discs and Jets at Super-Eddington Luminosity}
\author[T. Okuda, V. Teresi, E. Toscano and D. Molteni]{T. Okuda$^{1}$
\thanks{E-mail:okuda@cc.hokkyodai.ac.jp}, V. Teresi$^{2}$, E. Toscano$^{2}$, 
D. Molteni$^{2}$ \\
$^{1}$Hakodate College, Hokkaido University of Education, Hachiman-cho 1-2,
 Hakodate 040-8567, Japan\\
$^{2}$Dipartimento di Fisica e Tecnologie Relative, Universita di Palermo,
 Viale delle Scienza,  Palermo, 90128, Italy}
 
\begin{document}

\date{Accepted }

\pagerange{\pageref{firstpage}--\pageref{lastpage}} \pubyear{2004}

\maketitle

\label{firstpage}

\begin{abstract}
 Super-Eddington accretion discs with 3 and 15 $\dot M_{\rm E}$ around black 
holes with mass 10$M_{\odot}$ 
 are examined by two-dimensional radiation hydrodynamical calculations
  extending from the inner disc edge to $5 \times 10^4 r_{\rm g}$ and lasting
 up to $\sim 10^6 r_{\rm g}/c$.
The dominant radiation-pressure force in the inner region of the disc
accelerates the gas  vertically to the disc plane, and 
 jets with 0.2 -- 0.4$c$ are formed along the rotational axis.
In the case of the lower accretion rate, the initially anisotropic  
high-velocity jet expands outward and 
becomes gradually isotropic flow in the distant region.
The mass-outflow rate from the outer boundary is as large 
as $ \sim 10^{19}$ -- $10^{23}$ g s$^{-1}$, but it is variable 
and intermittent with time; that is, the outflow switches occasionally 
to inflow in the distant region.  
The luminosity also varies as $\sim 10^{40}$ --$10^{42}$
erg s$^{-1}$ on a long time-scale.   
On the other hand, the jet in the case of the higher accretion rate maintains 
its initial anisotropic shape even after it goes  far away.
 The mass-outflow rate and the luminosity attain to  steady values 
 of $3 \times 10^{19}$ g s$^{-1}$ and $ 1.3 \times 10^{40}$ erg s$^{-1}$, 
 respectively. 
 In accordance with the local analysis of the slim accretion disc model,
 the disc is thermally unstable in the case of 3 $\dot M_{\rm E}$ but stable in the case of 15 $\dot M_{\rm E}$.
 The super-Eddington model with 15 $\dot M_{\rm E}$ 
 is promising to explain  a small collimation degree of 
 the jet and a large mass-outflow rate observed in the X-ray source SS 433.

\end{abstract}


\section{Introduction}

 Disc accretion is an essential process for such dynamic phenomena as energetic
 X-ray sources, active galactic nuclei, and protostars.
 Since the early works by \citet{b23} and \citet{b25}, a great number 
  of papers have been devoted to studies of the disc accretion onto 
 gravitating objects, and 
 the standard accretion disc models with geometrically thin disc have 
 proven to be particularly 
 successful in applications to cataclysmic variables. 
 However, for very luminous accretion discs whose luminosity exceeds   
 the Eddington luminosity,    
 we need another disc model instead of the standard model
  because  the disc is 
 expected to have geometrically thick structure and  powerful outflows 
 from the disc.
 It is well known that the luminous accretion discs, where 
 the radiation-pressure and the electron-scattering are dominant, are thermally
 unstable if the usual $\alpha$-model for the viscosity is used. 
For the luminous disc, 
the slim accretion disk model 
 has been proposed by \citet{b1}, where strong advective cooling 
 depresses the thermal instability.  The slim disc model seems 
 to be reasonable for the accretion discs at the super-Eddington luminosity,
  but we need to reconfirm it from the point of view of a two-dimensional 
 simulation.
  The super-Eddington accretion discs were generally expected to
 possess vortex funnels and radiation-pressure driven jets from
 geometrically thick discs \citep{b25,b17,b7}.
 A two-dimensional hydrodynamical calculation of a super-Eddington 
  accretion disc around a black hole was first examined by \citet{b5,b6},
 and was discussed regarding SS 433.
 Although they showed a relativistic jet formation just outside a conical 
 photosphere 
 of the accretion disc and  a collimation angle of $\sim 30^{\circ}$ of the 
 jet,  the jet's mass flux is  too small and the collimation angle 
 is much larger than the one observed for SS 433.
 
\citet{b4} investigated an 
 adiabatic inflow--outflow solution of the accretion disc, where the 
 super-Eddington flow  leads to a powerful wind.
 \citet{b12} discussed the super-Eddington mass transfer 
  in binary systems and how to accrete in such super-Eddington accretion discs.
 Recent theoretical studies and two-dimensional
 simulations of the black holes accretion flows \citep{b10,b28,b18,b24,b27,b2}
 showed a new development 
 in the advection-dominated accretion flows (ADAFs) and the 
 convection-dominated accretion flows (CDAFs) which may be deeply related to
 the super-Eddington accretion flows.  
 Furthermore, from a point of view completely different from the 
 $\alpha$-viscosity models for accretion discs,
 recently developed rigorous works of global three-dimensional 
 magnetohydrodynamic (MHD) simulations of black hole accretion flows 
 should be noticed \citep{b8,b3}. They show that, without assumption of 
  the phenomenological $\alpha$-viscosity, magnetic fields play 
  the most important role in the angular momentum transport; that is, 
 accretion is driven  by turbulent stresses generated selfconsistently 
 by the magnetorotational instability. The results reveal a three component 
 structure: a warm Keplerian disc, a highly magnetized corona, and axial jet.
  
  \citet{b20} examined the super-Eddington black-hole model for SS 433 
  based on two-dimensional hydrodynamical calculations coupled 
 with radiation transport. Although, in this model, a high-velocity jet with 
 0.2 -- 0.4$c$ is formed  along the rotational axis,  
 a small collimation angle of 
 $\sim 0.1$ radian of the jets and a sufficient mass-outflow rate comparable 
 to $\sim 10^{19}$  g s$^{-1}$ could not be 
 obtained. Additionally, it was also suggested that this result might be a 
 transient behavior during a long disc evolution \citep{b30}.
 Keeping in mind these facts, we examine further super-Eddington accretion 
 discs and jets around the black hole extending from the inner disc edge to 
 $5 \times 10^4 r_{\rm g}$.

\section{ Model Equations}

 A set of relevant equations consists of six partial differential 
 equations for density, momentum, and thermal and radiation energy.
 These equations include the full viscous stress tensor, heating and cooling
 of the gas, and radiation transport.   
 The radiation transport is treated in the gray, flux-limited diffusion
 approximation \citep{b15}.
 We  use spherical polar coordinates ($r$,$\zeta$,$\varphi$), where $r$
 is the radial distance, $\zeta$ is the polar angle measured from 
 the equatorial plane of the disc, and $\varphi$ is the azimuthal angle.  
 The gas flow is assumed to be axisymmetric with respect to $Z$-axis 
 ($\partial /\partial \varphi=0 $) and the equatorial plane .
 In this coordinate system, the basic
 equations for mass, momentum, gas energy, and radiation energy 
 are written  in the following conservative form \citep{b13}:

 \begin{equation}
   { \partial\rho\over\partial t} + {\rm div}(\rho\mbox{\boldmath$v$}) =  0,  
 \end{equation}
 \begin{equation}   
  {\partial(\rho v)\over \partial t} +{\rm div}(\rho v \mbox{\boldmath$v$})  = 
   \rho\left[{w^2\over r} + {v_\varphi^2\over r}-{GM_* 
   \over (r-r_{\rm g})^2} \right] -{\partial p\over \partial r}+f_r +{\rm div}
   \mbox{\boldmath$S$}_r
   + {1\over r} S_{rr},  
 \end{equation}
 \begin{equation}  
  {{\partial(\rho rw)}\over \partial t} +{\rm div}(\rho rw\mbox{\boldmath$v$}) 
  = -\rho v_\varphi^2{\rm tan}\zeta-{\partial p\over\partial\zeta}
     +{\rm div}(r\mbox{\boldmath$S$}_\zeta)
    +S_{\varphi\varphi}{\rm tan}\zeta + f_\zeta , 
 \end{equation}
 \begin{equation}    
 {{\partial(\rho r{\rm cos}\zeta v_\varphi)}\over \partial t} 
     +{\rm div}(\rho r{\rm cos}
 \zeta v_\varphi\mbox{\boldmath$v$}) = 
 {\rm div}(r {\rm cos}\zeta \mbox{\boldmath$S$}_\varphi),
\end{equation}
\begin{equation}  
  {{\partial \rho\varepsilon}\over \partial t}+
    {\rm div}(\rho\varepsilon\mbox{\boldmath$v$})
      = -p\;\rm div \mbox{\boldmath$v$} + \Phi - \Lambda, 
\end{equation}
and
\begin{equation}       
  {{\partial E_0}\over \partial t}+ {\rm div}\mbox{\boldmath$F_0$} +
        {\rm div}(\mbox{\boldmath$v$}E_0 +\mbox{\boldmath$v$}\cdot P_0) 
        = \Lambda 
      - \rho{(\kappa +\sigma)\over c}\mbox{\boldmath$v$}\cdot
      \mbox{\boldmath$F_0$} ,
 \end{equation} 
 where $\rho$ is the density, $\mbox{\boldmath$v$}=(v, w, v_\varphi)$ are the
 three velocity components, $G$ is the gravitational constant,
 $M_*$ is the central mass, $p$ is the  gas pressure,
 $\varepsilon$ is the specific internal energy of the gas,  $E_0$ is 
 the radiation energy density per unit volume, and $P_0$ is the radiative
  stress tensor. The subscript ``0'' denotes
  the value in the comoving frame and that the equations are correct
   to the first order of $\mbox{\boldmath$v$}/c$ \citep{b11}.
 We adopt the pseudo-Newtonian potential \citep{b21}
 in equation (2), where $r_{\rm g}$ is the Schwarzschild radius.
 The force density $\mbox{\boldmath$f$}_{\rm R}=(f_r,f_\zeta)$ exerted 
 by the radiation field is given by
\begin{equation} 
  \mbox{\boldmath$f$}_{\rm R}=\rho\frac{\kappa+\sigma}{c}\mbox{\boldmath$F_0$}, 
\end{equation} 
 where $\kappa$ and $\sigma$ denote the absorption and scattering 
 coefficients and $\mbox{\boldmath$F_0$}$ is the radiative flux 
 in the comoving frame.
 For the opacities $\kappa$ we use  polynomial fits to the Rosseland mean opacities
  given by \citet{b16}.
 $S=( \mbox{\boldmath$S$}_r,\mbox{\boldmath$S$}_\zeta,\mbox{\boldmath$S$}_
 \varphi)$ denote the viscous stress tensor which includes all components due
 to the three velocity components \citep{b19}. 
 $\Phi=(S\;\nabla)\mbox{\boldmath$v$}$ is the viscous dissipation 
 rate per unit mass.

 The quantity $\Lambda$ describes the cooling and heating of the gas, 
 
 \begin{equation}      
      \Lambda = \rho c \kappa(S_*-E_0), 
\end{equation}
 where $S_*$ is the source function and $c$ is the speed of light. 
 For this source function, we assume local thermal equilibrium $S_*=aT^4$, 
 where $T$ is 
 the gas temperature and $a$ is the radiation constant.
 For the equation of state, the gas pressure is given by the ideal gas law, 
 $p=R_{\rm G}\rho T/\mu$, where $\mu$ is the mean molecular weight 
 and $R_{\rm G}$ is the gas constant. 
  The temperature $T$ is proportional to the specific
 internal energy, $\varepsilon$, by the relation $p=(\gamma-1)\rho\varepsilon
  =R_{\rm G}\rho T/\mu$, where $\gamma$ is the specific heat ratio.  
  To close the system of 
 equations, we use the flux-limited diffusion approximation \citep{b15} 
 for the radiative flux:
\begin{equation}
   \mbox{\boldmath$F_0$}= -{\lambda c\over \rho(\kappa+\sigma)}
   {\rm grad}\;E_0, 
\end{equation}

\noindent and
\begin{equation}
   P_0 = E_0 \cdot T_{\rm Edd}, 
\end{equation}
where  $\lambda$ and $T_{\rm Edd}$ are the {\it flux-limiter} and the 
 {\it Eddington Tensor}, respectively, for which we use the approximate
 formulas given in \citet{b13}.
 The formulas fulfill the correct
 limiting conditions in the optically thick diffusion limit and the
 optically thin streaming limit, respectively.
 
 For the kinematic viscosity, $\nu$, we adopt a 
 modified version \citep{b22}
  of the standard $\alpha$- model.
 The modified prescription for $\nu$ is given by
\begin{equation}
  \nu=\alpha\; c_{\rm s}\; {\rm min}\left[H_{\rm p},H\right], 
\end{equation}
 where $\alpha$ is a dimensionless parameter,
 $c_{\rm s}$  the local sound speed,  and $H$ the disc height. 
 $H_{\rm p}=p /\mid {\rm grad} \;p\mid$  is the mid-plane pressure
 scale height inside of the disc but the local pressure scale height outside of the disc.
 
\section{Numerical Methods}

 The set of partial differential equations (1)--(6) is 
 numerically solved by a finite-difference method under adequate initial 
 and boundary conditions.
 The numerical schemes used are basically the same as that described by 
  \citet{b13} and \citet{b19}. 
 The methods are based on an explicit-implicit finite difference scheme.
  Grid points in the radial direction are spaced logarithmically
  as $\Delta r/r = 0.1$,
 while grid points in the angular direction are equally 
 spaced, but more refined near the equatorial plane, typically $\Delta \zeta=
 \pi/150$ for $\pi/2 \geq \zeta \geq \pi/6 $ and $\Delta \zeta=
 \pi/300$ for $\pi/6 \geq \zeta \geq 0 $.
 Although the radial mesh-sizes do not have a fine 
 resolution to examine detailed disc structure, we consider the mesh-sizes to
 be sufficient for examination of 
 the global behaviors of the disc, the jet, and the mass-outflow rate, as is 
 later shown in subsection 4.2.

 \subsection{Model Parameters}
 We consider a Schwarzschild black hole with mass $M_*= 10 M_{\odot}$
 and take the inner-boundary radius $R_{\rm in}$ of the computational domain
 as $2 r_{\rm g}$.
 The model parameters used
  are listed in table 1, where $R_{\rm max}$ is the
  outer boundary radius, and $\dot m$ is
 the input accretion rate normalized to the Eddington critical accretion rate
 $\dot M_{\rm E}$ which is given by 
 
 \begin{equation}
 \dot M_{\rm E}   = 16 L_{\rm E} / c^2,
 \end{equation}
 where $L_{\rm E}$ is the Eddington luminosity.
 $L$,  $\dot M_{\rm out}$, and $\dot M_{\rm in}$ in table 1 denote the total
 luminosity through the outer boundary surface, the mass-outflow rate ejected from the outer 
 boundary, and the mass-inflow rate swallowed into the black hole through 
  the inner boundary  at the final  state.
 The viscosity parameter $\alpha = 10^{-3}$ is adopted, since in the previous 
 examination we found that the low value of $\alpha=10^{-3}$ is more 
 advantageous to formation of the collimated high velocity jets than the high 
 value case with $\alpha$ = 0.1 \citep{b20}.

\begin{table*}
\centering
\caption{Model parameters}
\begin{tabular}{@{}ccccccc} \hline \hline
Model & $\dot m$  & $\dot M (\rm{\;g \;s^{-1}})$ & $R_{\rm max}/r_{\rm g} $ &
 $L/ L_{\rm E}$ & $\dot M_{\rm out}/\dot M$ &  $\dot M_{\rm in}/\dot M$ \\ \hline
 1    &  3  &  8 $\times 10^{19} $ & $ 5 \times 10^4$ & variable & variable & 0.4 \\  
 2   &  15  &  4 $\times 10^{20} $ & $ 5 \times 10^4$ & 8 & 0.08 & 0.2
   \\ \hline
\end{tabular}
\end{table*}

\subsection
{ Initial Conditions}
 The initial conditions  consist of a cold, 
 dense, and optically thick disc and a hot, rarefied, and optically
 thin atmosphere around the disc. 
 The initial disc at $r/r_{\rm g} \ge 3$ is approximated by the 
 Shakura-Sunyaev  standard model but, at 2 $\leq r/r_{\rm g} \le 3$,
 it is taken to be sub-Keplerian disc with a free-fall velocity at the inner
  boundary.
 To examine how the high velocity jet evolves through the surrounding matter,
  it is important, for us, to set up the appropriate density distribution
   around the disk.
  Observations and hydrodynamical modeling of SS 433 jet \citep{b14} suggest 
  that the X-ray emitting region of SS 433 would be in the range of $r = 
  10^{10}$ -- $10^{12}$ cm from the central source and have gas densities 
  of $\sim 10^{-12}$ -- $10^{-10}$ g cm$^{-3}$ at the base of the X-ray jets. 
 Focussing on these facts,  we take the density at the outer 
 boundary radius of $1.5 \times 10^{11}$ cm to be $\sim 10^{-14}$ g cm$^{-3}$.
  Furthermore, assuming the density distribution of $\rho \sim r^{-1}$, 
  we construct an initial hot rarefied 
  atmosphere around the disc to be approximately in radially hydrostatic 
  equilibrium.

 \subsection
 {Boundary Conditions}

   Physical variables at the inner boundary, except for the velocities,
    are given by extrapolation of the variables near the boundary. 
  However, we impose limited conditions that the radial velocities are given 
  by a free-fall velocity and the angular velocities are zero.
   On the rotational axis and the equatorial plane, 
 the meridional tangential velocity $w$ 
 is zero and all scalar variables must be symmetric relative to these axes.
 The outer boundary at $r=R_{\rm max}$ is divided into two parts. 
 One is the disc boundary through which matter is entering from 
 the outer disc.
 At the outer-disc boundary we assume a continuous inflow of matter 
 with a constant accretion rate $\dot M$.   
 The other is the outer boundary region above the
 accretion disc. We impose free-floating conditions on this outer boundary 
 and allow for outflow of matter, whereas any inflow is prohibited here. 
 We also assume the outer boundary region above the disc to be 
 in the optically-thin limit, 
  $\vert \mbox{\boldmath$F_0$} \vert \rightarrow c E_0$.

       \begin{figure}
       \includegraphics[width=140mm,height=80mm]{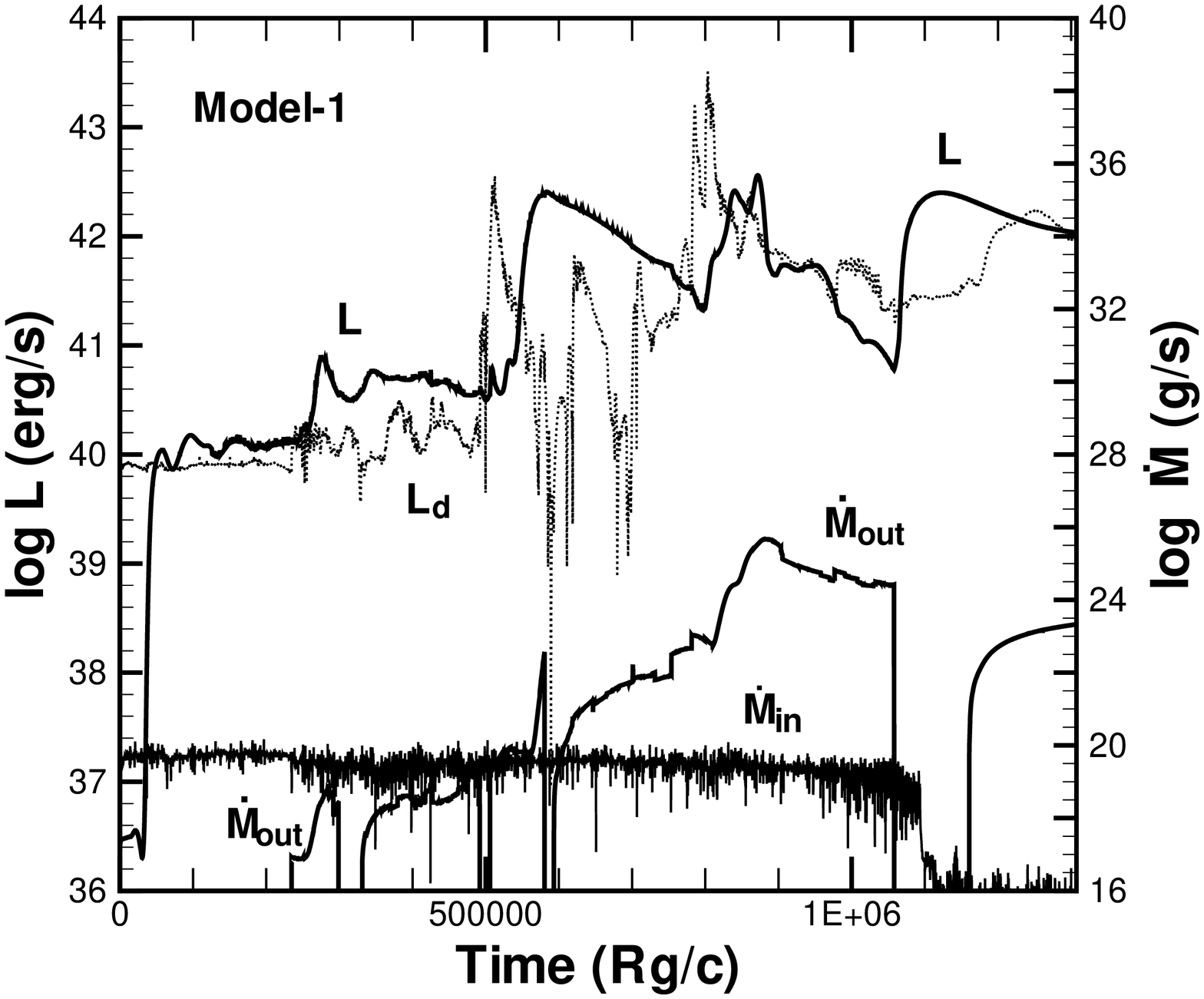}
       \caption{Time evolution of total luminosity $L$ emitted from the outer
        boundary surface, disc luminosity $L_{\rm d}$ from the disc surface,
        mass-outflow rate $\dot M_{\rm out}$ from the outer boundary,
        and mass-inflow rate $\dot M_{\rm in}$
        swallowed into the black hole through the inner boundary 
        for model 1, where time is 
       shown in units of $r_{\rm g}/c$.}
      \label{fig1}
      \end{figure}

     \begin{figure}
       \includegraphics[width=140mm,height=110mm]{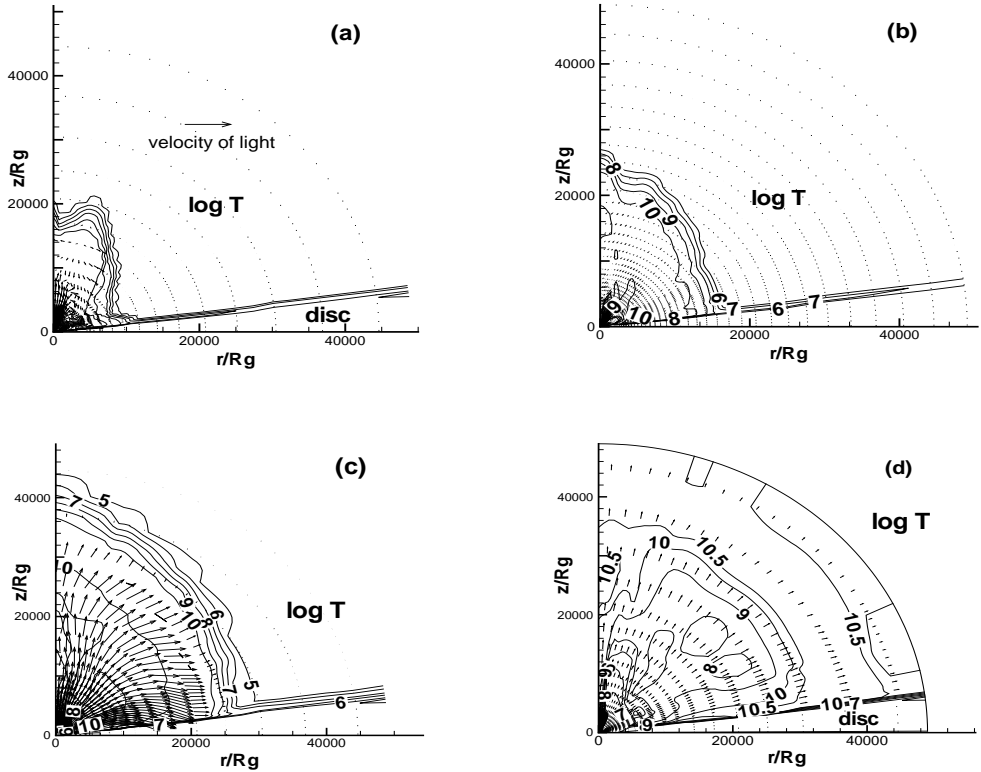}
       \caption{Velocity vectors and temperature contours in logarithmic
        scale on the meridional plane at $t = 6 \times 10^4$ (a), $1.5 \times 
        10^5$ (b), $2.5 \times 10^5$ (c), $2.8 \times 10^5 r_{\rm g}/c $ (d)  
        for model 1. 
        The contour lines with labels of log $T$ = 5, 6, 7, 8, 
        9, 10, and 10.5 are shown. 
        The reference vector of light is shown by a long arrow in (a).
        The initially anisotropic high-velocity jet (a) generated in the 
        inner region of the disc evolves into the isotropic outflow (d) in 
        the distant region. `disc' in the figure denotes the cold and dense 
        accretion disc.}        
      \label{fig2}
      \end{figure}

\section{Numerical Results}

 \subsection{Case of $\dot M =  3  \dot M_{\rm E}$}
 
  Model 1 with $\dot M =3  \dot M_{\rm E}$  has the same input accretion
 rate as  BH-1 in the previous disc model \citep{b20}, but the computational 
 domain used is by two orders of magnitude larger than the previous one.
 This is due to the following reasons.
   We use the initial discs based on 
    the Shakura-Sunyaev  model. In the super-Eddington 
    accretion regime, the initial conditions should be carefully used
    on treatment of the outer disc-boundary, 
   because we don't know a priori reasonable conditions of the very luminous 
   and thick disc.
   However, if  the outer boundary radius is taken to be considerably large, 
   this problem will be avoidable because the outermost disc is 
    well described by the Shakura-Sunyaev  model.
   Another reason is due to a requirement that we want to examine the 
   behavior of high-velocity jet over a distant region, which are 
   observed in the X-ray source SS 433.

   The initial disc used is geometrically thick as $H/r$  
   $\sim 1$ in the inner region,
   but the relative disc height $H/r$ and 
   the ratio  $\beta$ of the gas pressure to the total pressure are $\sim$ 
   0.05 -- 0.1  and $\sim$ 0.95 -- 0.98, respectively,
   at the outer disc boundary for models 1 and 2.

       \begin{figure}
       \includegraphics[width=140mm,height=110mm]{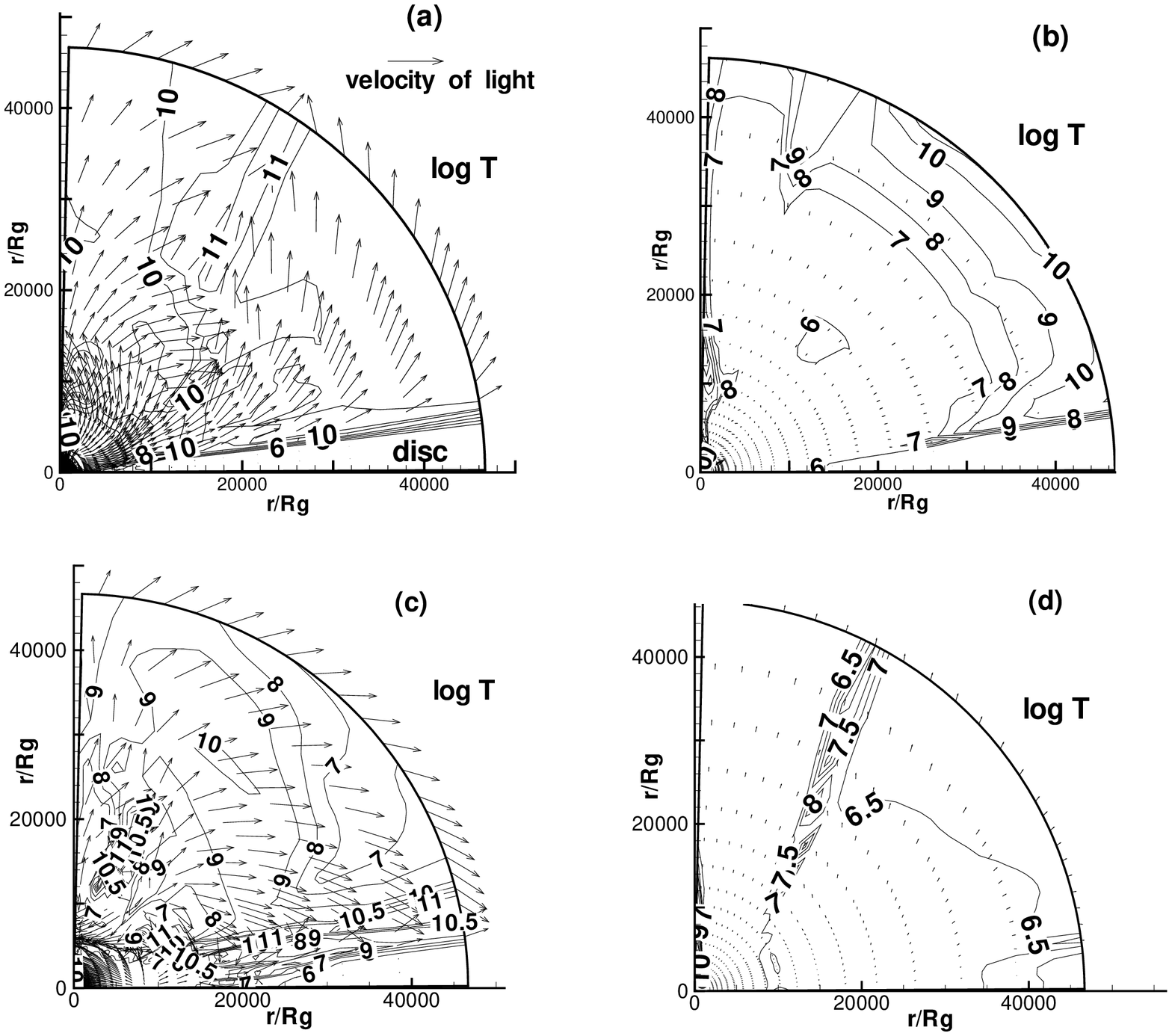}
        \caption{Same as figure 2 but at $t = 5.1 \times 10^5$ (a), $ 5.9 
      \times 10^5$ (b), $8.0 \times 10^5$ (c), $1.1\times 10^6$ (d)  
       $r_{\rm g}/c $ for model 1.  
       (a) and (c) show active outflows in the high temperature state, but (b)
        and (d) show inflows (no mass-outflow) in the low temperature state.
                 }
       
      \label{fig3}
      \end{figure}
 
 Fig.~1 shows the time evolutions of the total luminosity $L$ emitted from the 
  outer boundary surface, the disc luminosity $L_{\rm d}$ emitted from the disc
   surface, the mass-outflow rate $\dot M_{\rm out}$ from the outer boundary,
  and the mass-inflow rate  $\dot M_{\rm in}$ swallowed into the black
   hole through the inner boundary for model 1. 
 From the luminosity curve, we can see three characteristic stages:  stage 1
 ( $t$ = 0 -- $2.5\times 10^5 r_{\rm g}/c $ ),  stage 2 ( $t =  2.5$ -- $5.5\times 
 10^5 r_{\rm g}/c $ ), and  stage 3 ( $t \ge 5.5\times 10^5 r_{\rm g}/c $ ).
 At the stage 1, matter in the inner region of the disc is ejected strongly 
 outward and propagates vertically to the disc plane. 
 The high-velocity jet with a velocity of 
 $\sim 0.2c$  reaches the outer boundary in the polar direction at $t \sim 
 R_{\rm max}/0.2c \sim 2.5 \times 10^5 r_{\rm g}/c $, and  then the 
 mass outflow through the outer boundary begins, and the total luminosity $L$ 
 attains to $\sim 10^{40}$ erg s$^{-1}$.
  At the stage 2, the luminosity flares up to $\sim 4\times 10^{40}$ erg 
  s$^{-1}$
  and maintains its high value for a time of $3 \times 
  10^{5}r_{\rm g}/c $ and switches to the stage 3, flaring up by more than
 one order of magnitude.  During the stage 3, the luminosity shows 
 quasi-periodic variations with a period of $ 2.5\times 
 10^{5}r_{\rm g}/c $ and an amplitude of factor ten.
 Through these stages, the mass-outflow rate $\dot M_{\rm out}$ 
  becomes as high as $\sim 10^{19}$
 -- $10^{23}$ g s$^{-1}$, but it is variable and intermittent as is 
 found in Fig.1; that is, the outflow switches occasionally to inflow.
 The mass-inflow rate $\dot M_{\rm in}$ fluctuates
 rapidly, but the averaged rate is constantly $\sim 3 \times 10^{19}$ g s$^{-1}$ 
 except the final phases, 
 where $\dot M_{\rm in}$ becomes very small as  
 $\sim 10^{16}$ g s$^{-1}$. In model 1, the heating wave (see Fig.~7) 
 and the density wave, which generate initially in the innermost disc region, 
 propagate outward  with 
 times and reach the outer disc region of $r \sim 3 \times 10^4 r_{\rm g} $ 
  at $t \sim 1.1 \times 10^6 r_{\rm g}/c $. At these phases, a part of the waves
  begins to back inward as an ingoing wave, and gradually the ingoing density wave 
   leads to lower density in the inner disc region and resultantly to smaller
   mass-inflow rate at the inner boundary.

  $L$ and $L_{\rm d}$  are given by  calculating $\int {\bf F_0} d{\bf S}$, 
  where the surface integral is taken over the outer boundary surface and
  the disc surface, respectively.
 Although it is difficult for us to specify correctly the disc  
 surface in the geometrically thick  disc, we define it here as a 
 location where the density drops to a tenth of the central density 
 on the equatorial plane.
 This may lead to some errors in estimation of the disc surface and
 accordingly $L_{\rm d}$.  The complicated features of $L_{\rm d}$ during 
 the stage 3 in Fig.1  may be due to unreliable determinations of 
 the disc surface.
 If the atmosphere above the disc is optically thin,
 $L_{\rm d}$  must agree with $L$. $L_{\rm d}$ has an initial value of $\sim
  8 \times 10^{39}$ erg s$^{-1}$ derived from the Shakura-Sunyaev model, 
 while  $L$ is 
 initially  small and becomes comparable to $L_{\rm d}$ at the phase $ t
  \sim R_{\rm max}/c$ = $5 \times 10^4 r_{\rm g}/c $ when radiation from the
 disc attains to the outer boundary radius. 
 After $t \sim 2.5 \times 10^5 r_{\rm g}/c$, the disc luminosity $L_{\rm d}$ 
 arises to $ \sim 1.3 \times 10^{40}$ erg s$^{-1}$ and fluctuates around 
 the value by a factor 2 during a time of $2.5 \times 10^5 r_{\rm g}/c$,
 and then increases sharply  to $3 \times 10^{42}$ erg s$^{-1}$.  
 The total luminosity $L$
  also denotes same behavior as the disc luminosity but with
  a phase lag of $5 \times 10^4 r_{\rm g}/c$, which is the radiation transit 
  time  from the disk surface to the outer boundary radius.  
  These time-dependent behaviors
  of the luminosities are discussed later in terms of the thermal instability
   of the slim disc.
 
 Fig.~2 denotes the temperature contours with velocity vectors at $t = 6 
  \times 10^4$,  $1.5 \times 10^5$,  $2.5 \times 10^5$, and $ 2.8 \times 10^5 r_{\rm g}/c$ 
  in the stage 1 for model 1.
   The evolutionary features of outward moving jets are found here,
   where the outermost crowded contours of temperature show the 
   shocked region due to the 
  interaction of the jet with the surrounding gas, and the reference velocity 
  of light is indicated by a long arrow in Fig.2 (a). 
   The temperatures are as high as $\sim 10^8$ -- $10^{10}$ K in 
   the high-velocity region  and `disc' in the figure denotes the cold and 
   dense disc.
   The initially anisotropic jet (a) propagates vertically to the disc
   plane but gradually loses its anisotropic nature with increasing
   times, and finally the jet becomes isotropic flow (d) after the jet reaches 
   the outer boundary. The figure also shows that there exist 
   an inactive phase (b) and an active phase (c) of outflows. 
   Fig.~3 shows the temperature contours with velocity vectors at $t = 5.0 
  \times 10^5$,  $5.9 \times 10^5$,  $8.0 \times 10^5$, and $1.1 \times 10^6
   r_{\rm g}/c$ in the stage 3. Fig.~3 (a) and (c) show powerful outflows 
  in the high-temperature
  state above the disc, whereas Fig.~3 (b) and (d) denote inflow states 
  in the low-temperature state. The inactive inflow states, accompanying with
   no mass-loss, are likely to occur after a large flare up of the luminosity.

    \begin{figure}
       \includegraphics[width=120mm,height=80mm]{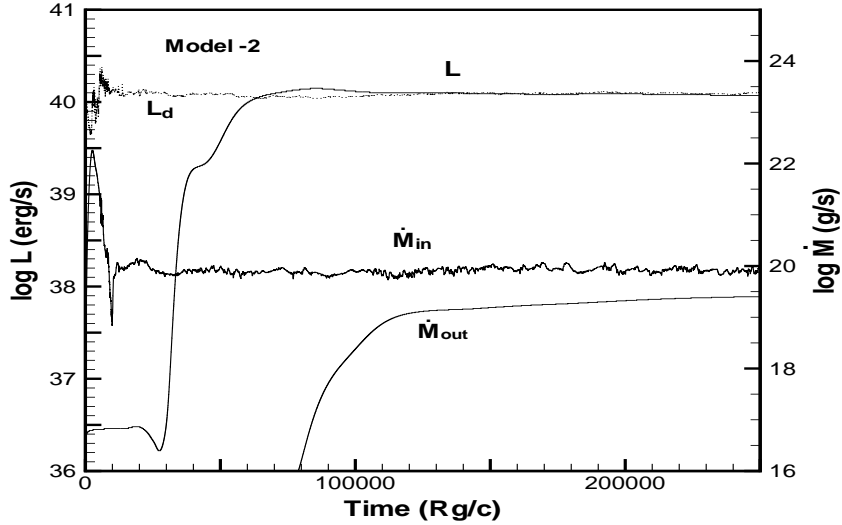}
       \caption{Time evolution of total luminosity $L$, disc luminosity 
       $L_{\rm d}$, mass-outflow rate $\dot M_{\rm out}$ from the outer 
        boundary, and mass-inflow rate $\dot M_{\rm in}$
        swallowed into the black hole through the inner boundary 
        for model 2.}
      \label{fig4}
      \end{figure}

       \begin{figure}
       \includegraphics[width=140mm,height=110mm]{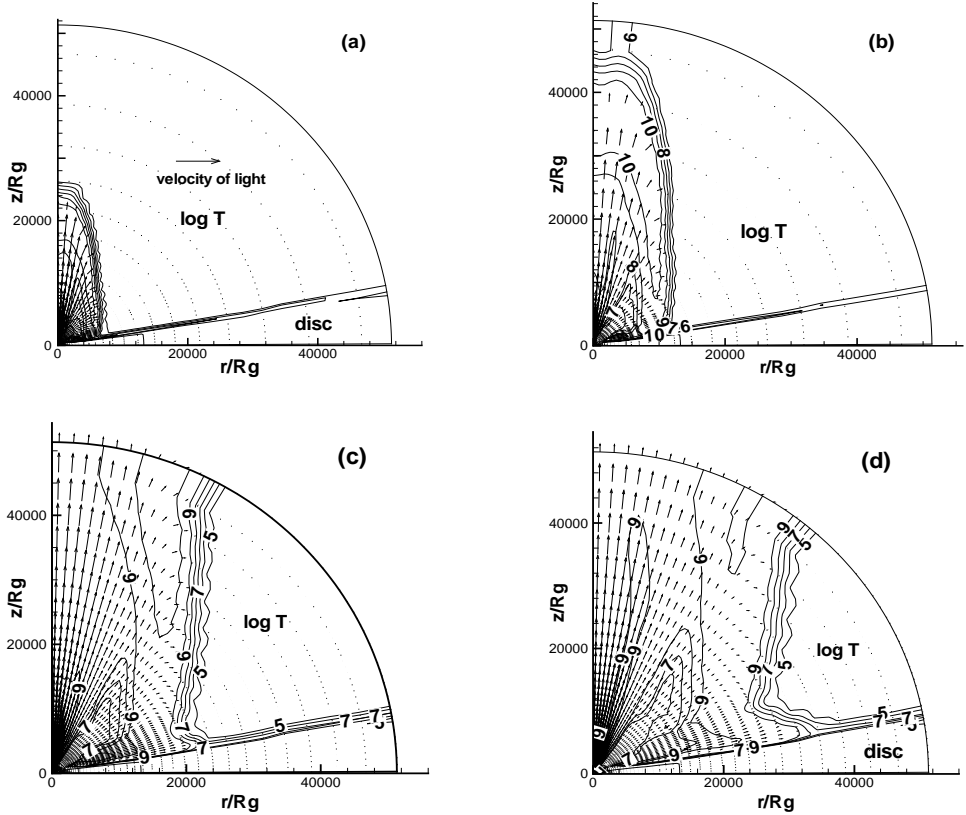}
       \caption{Velocity vectors and temperature contours in logarithmic scale 
       on the meridional plane at $t = 5 \times 10^4$  (a), $8 \times 
        10^4$  (b), $1.5 \times 10^5$  (c), $2.5 \times 10^5 $  (d)  but for 
        model 2.
       Differently from model 1, the initially anisotropic high-velocity
       jet along the rotational axis maintains its anisotropic shape
        when it expands outward.
      }
      \label{fig5}
      \end{figure}

 As is found in ~Figs. 2 and 3, the gas temperatures in the outflow region 
 outside of the disc are partly too high as $\sim 10^9$ -- $10^{11}$ K.
 These temperatures may be unreliable because we did not take account of other 
 physical processes, such as Compton processes and pair production-annihilation, which would be important at such high temperatures.  As far as the disc is
 concerned, however, the temperatures are in the range of 
 $10^6 \leq T < 10^8$ K.

 \subsection{Case of $\dot M= 15 \dot M_{\rm E}$}
  The time evolutions of $L, L_{\rm d}, \dot M_{\rm out}$, and 
 $\dot M_{\rm in}$ for model 2 are shown in Fig.~4.
  At the initial phases, the disc luminosity $L_{\rm d}$ and 
  the mass-inflow rate $\dot M_{\rm in}$ are very dependent on the initial 
  structure of the inner disc and they fluctuate largely, but they settle 
  to their steady state values in a time-scale of $ \sim  2 \times 10^4 
  r_{\rm g}/c $.  
  The total luminosity $L$  becomes comparable to the disc
  luminosity $L_{\rm d}$ at  $t \sim R_{\rm max}/c = 5 \times 10^4 
  r_{\rm g}/c$ when radiation from the disc arrives at the outer boundary,
  and finally it accords with $L_{\rm d}$.
  On the other hand, after $t \sim R_{\rm max}/0.4c = 10^5 r_{\rm g}/c $
  which is the jet transit time from the disc to the outer boundary, 
  the mass outflow 
  begins and $\dot M_{\rm out}$ gradually settles down to a steady state 
  value of $3 \times 10^{19}$ g s$^{-1}$. Finally, $ L = L_{\rm d} = 1.3 
  \times 10^{\rm 40}$ erg s$^{-1}$. The mass-inflow rate swallowed into
   the black hole fluctuates always by a small factor around $ 8 \times 
   10^{19}$ g s$^{-1}$.
  
     \begin{figure}
       \includegraphics[width=140mm,height=80mm]{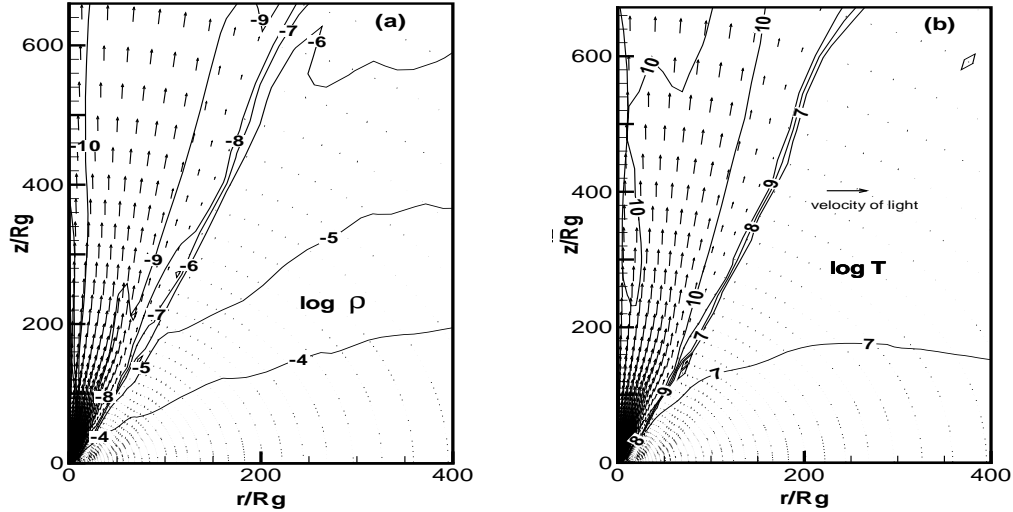}
       \caption{Velocity vectors and contours of density (a)
        and temperature (b)
       in the inner region of the disc and the high-velocity jet
        at $t=  2.5 \times 10^5$ for model 2. The contour labels denote the
         logarithmic values of the density $\rho$ (g cm$^{-3}$) and the
         temperature $T$ (K).}
      \label{fig6}
      \end{figure}

  Fig.~5 shows the evolutionary features of the high-velocity jet at
  $t = 5 \times 10^4$ (a), $8 \times 10^4$ (b), $1.5 \times 10^5$ (c), 
   and $2.5 \times 10^5$ (d).
  The high-velocity jet propagates vertically
  to the disc plane in the same way as model 1.
  The jet expands gradually far from the rotational axis with 
  increasing times.
  After the phase in Fig.~5 (b), the jet arrives at the outer boundary 
  in the polar direction but the anisotropic nature of the
  jet is remained even after it passed through the outer boundary.
  If shapes of the jets found in Figs.~5 (a) and (b) are always kept 
  throughout the time evolution,
  we can expect a small collimation angle of $\le 10^{\circ}$ for the jet.

  Fig.~6 shows the velocity vectors and the contours of density $\rho$ 
  (g cm$^{-3}$ ) and temperature $T$ (K)  of the disc
   and the high-velocity jet  in the inner region at $t= 2.5 \times 10^5 
   r_{\rm g}/c$ for model 2. This shows  a rarefied, hot, and optically 
   thin high velocity jet region and a dense, cold, and geometrically thick 
   disc region. The high-velocity region and the 
   disc region are bounded by the funnel wall, a barrier where the effective 
   potential due to the gravitational potential and the centrifugal one 
   vanishes. The jet velocity  in the funnel region between the funnel
   wall and the Z-axis amounts to 
   $\sim 0.3c$. In the inner disc, the flow is convectively unstable. 
  In the innermost region of the disc, there exist 
  the inner advection-dominated zone and the outer convection-dominated zone.
  The transition radius $r_{\rm tr}$ between the two zones is approximately 10 
  $r_{\rm g}$.  At $r \geq r_{\rm tr}$, roughly half the mass at any time 
  at any radius is flowing in and flowing out, respectively, and the 
  mass-inflow rate balances the 
  mass-outflow rate. Near to the transition region, by way of the convective
 zones and the accretion zone, matter is accreted towards the equatorial plane.
 The accreting matter, which is carried to the transition by convection, 
  partly diverts into the high-velocity region and partly flows into the inner
  advection-dominated zone.  The gas diverged into the optically thin 
  high-velocity region is originally subsonic but is soon accelerated by 
 the strong radiation field up to a relativistic velocity. 
 These features of the accreting and outflow materials in the inner region are
 also found in model 1, even when it is at an inflow phase in the outer region.

  In Fig.~6, we see that at the disc boundary the densities lie 
  between label $-$ 6 and $-$ 7 and  a temperature of $\sim 10^7$ K in 
 the disc jumps to  $\sim 10^8$ K. 
 The opening angle of the high-velocity jet to
 the Z-axis is $\sim 30^{\circ}$.
  These density and temperature profiles, the jet velocity,  and the opening 
 angle  show good agreements 
 with those in the previous calculation extending to $400 r_{\rm g}$ \citep{b20}, 
 where the relative radial mesh of $\Delta r/r$ = 0.03 is finer than 0.1 in 
 the present one but the mesh-sizes in the angular direction have same
 resolution.
 Therefore, the present results over an extended region of  
 $\sim 5 \times 10^4 r_{\rm g}$ are 
  considered to be also valid in spite of the coarse radial mesh-sizes.

   \begin{figure}
       \includegraphics[width=120mm,height=80mm]{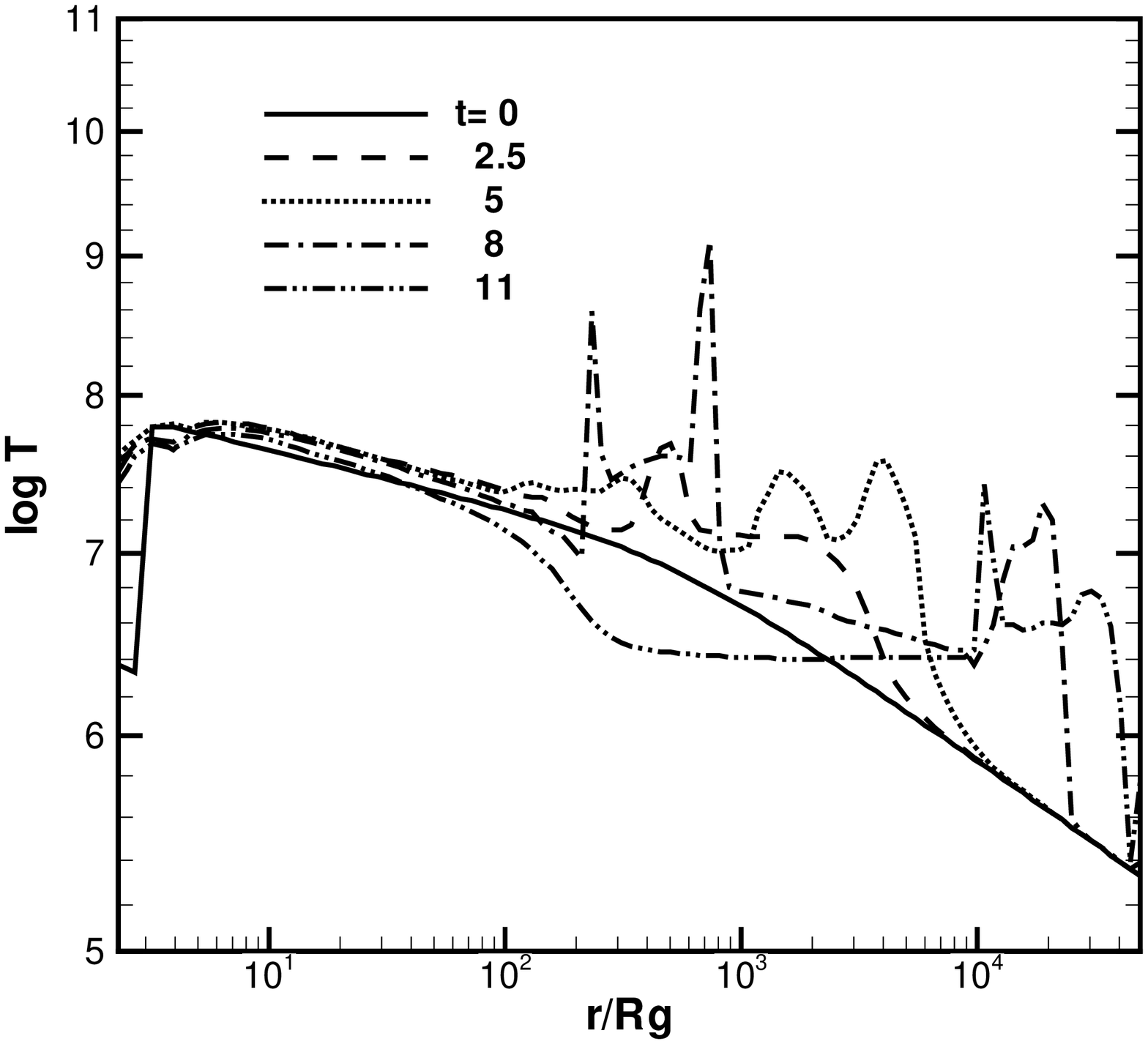}
       \caption{Temperature versus radius on the equatorial plane 
        at $t$= 0 (line), 
       $ 2.5 \times 10^5$ (dashed line), $5.1 \times 10^5$ (dotted line), 
       $8.0 \times 10^5$ (dash-dot line), and  $1.1\times 10^6$ (dash-dot-dot line)
        $r_{\rm g}/c $ for model 1.}
      \label{fig7}
      \end{figure}
   \begin{figure}
       \includegraphics[width=120mm,height=80mm]{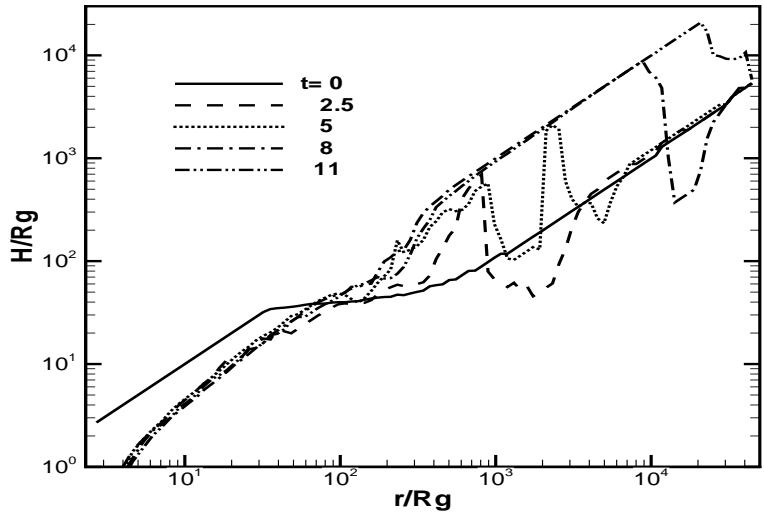}
       \caption{Disc height $H$ versus radius on the equatorial plane at 
       $t$= 0 (line), 
       $ 2.5\times 10^5$ (dashed line), $5.1 \times 10^5$ (dotted line), 
       $8.0\times 10^5$ (dash-dot line), and  $1.1\times 10^6$
         (dash-dot-dot line) $r_{\rm g}/c $ for model 1. }        
      \label{fig8}
      \end{figure}

   The disc and the jet in model 2  attain  almost  to their steady
   states, while model 1 shows variable luminosities and non-steady gas flows.
   These behaviors of the disc 
   luminosity in models 1 and 2 are considered in terms of the thermal 
   instability of the slim accretion disc model.  
   The thermal instability of the standard disc is 
   interpreted by the relation between the accretion rate $\dot m$ and
   the surface density $\Sigma$ at a fixed radius $r$.  
   When the curve of $\dot m = \dot m(\Sigma)$ has a characteristic S-shape 
   with three branches (lower, middle, and upper),
   the discs in the upper and lower branches are stable against thermal 
   instability, but in the middle branch it is unstable.
   If the disc is under the unstable middle branch, it is expected that
   the disc exhibits a limit-cycle behavior of luminosity.
   
   Taking account of advective cooling, \cite{b1} found the S-shaped $
   \dot m - \Sigma$ curve at $r=5r_{\rm g}$, corresponding to moderately 
   super-Eddington 
   accretion rates from the transonic solutions of a black hole
   with $M_*=10 M_{\rm \odot}$ and $\alpha=10^{-3}$.
   According to their results,
   the inner disc with $\dot m=15$ in our model 2 corresponds to the  
   stable upper branch, while the disc with $\dot m=3$ belongs marginally 
  to the unstable
   middle branch and it locates near the turning
   point between the upper and middle branches.
   Fig.~7 denotes the temperature versus radius on the equatorial plane 
   at $t=0$,  $2.5\times 10^5$,  $5.1 \times 10^5$,  $8.0 \times 10^5$, 
   and $1.1\times 10^6 r_{\rm g}/c$ for model 1. 
   The maximum luminosity appears at $t= 5.1\times
   10^5 r_{\rm g}/c$.  The evolution of the disc temperature is qualitatively 
     similar to the results by \citet{b9} and \citet{b29}, who found a 
   limit-cycle behavior of the slim disc with $\dot m =0.06$ and 
   $\alpha=0.1$ by one-dimensional calculations of vertically integrated 
   equations.
    The model with $ \dot m=0.06$ and $\alpha =0.1$, which belongs to the 
    unstable middle branch of $\dot m$ -- $\Sigma$ curve at $r/r_{\rm g}=5$, 
    locates just near another turning point between the unstable middle 
    branch and stable lower branches.
   In their results, the outgoing heating wave reaches about 100$r_{\rm g}$,
   while in ours the heating wave attains to  $\sim 3 \times 10^4 r_{\rm g}$.
   The disc height $H$ versus radius at same phases as Fig.~7 
   is shown in Fig.~8.
   In the initial phase, $H/r$ is $\sim 0.1$ at $r/r_{\rm g} \ge 10^{3}$,
  but the outgoing heating wave leads to $H/r \sim 1$ in the outer disc
  region and the radiation pressure is dominant there.
   Compared with the results by \citet{b9} and \citet{b29}, the heating wave 
   reaches the extensively distant region because of the much higher input 
   accretion rate used.  
   Apart from the effects due to the much higher accretion rate, we can see 
    similar characteristics of the disc evolution between theirs and us and
    expect a limit cycle behavior of the luminosity in model 1.
   However, in order to confirm the limit cycle phenomenon of the luminosity,
   the computational time in model 1 may not be sufficient because the 
  temperature evolution shows only the phases of the outgoing heating wave.

   On the other hand, the super-Eddington
   disc with $\dot m=15$ is  stable against the thermal instability
    due to the strong advective cooling.  
   In this model, the outgoing heating wave with temperatures of $10^8$ K $
    \leq T \leq 10^{10}$ K generates near the inner edge of the disc 
    at the initial phases and reach about $\sim 10^2 r_{\rm g}$.
   However, these waves  fade out soon and successively the second heating wave
   generates near the inner edge and oscillates at $2 \leq r/r_{\rm g} 
   \simeq 10$ during a time of  $ \sim 10^4 r_{\rm g}/c$, and finally they are
   stabilized completely.

\section{Comparison with SS 433}
  SS 433 is a typical stellar black-hole candidate at a highly super-Eddington 
  luminosity.
  Our results may be compared with the observations in SS 433.
  The observations show many emission lines of heavy elements, such
  as Fe and Ni, which denote the unique red and blue Doppler shifts of 0.26 $c$
  in the X-ray region. 
  The collimation angle of SS 433 jets seems to be as small 
  as 0.1 radian (several degrees).
 From an observational constraint of the mass-outflow rate, 
  \citet{b14} shows  $\dot M_{\rm out} \ge 3 \times 10^{19}$ g s$^{-1}$
 in SS 433. 
  
  The unique velocity, 0.26 $c$, may be reasonably explained in terms of 
 the relativistic velocities accelerated by the radiation-pressure force in the
 inner region of the super-Eddington disc.
 In our models, the collimation degree of the high velocity jets will be
  dependent on the initial density distribution of the matter 
  around the disc.  However, the density distribution used is not a special 
  type of profile but a usual spherical one with $\rho \propto r^{-1}$. 
  In model 1, 
  the initial jet is strongly collimated  in the polar region, but it becomes 
  isotropic flow in the distant region of $\sim 10^{11}$ cm, where the X-ray
  jet is observed in SS 433.  While in model 2 the small collimation angle 
  of the jet is  maintained throughout the jet propagation.
  If the anisotropic nature of the jet found in Fig.~5 (a) and (b)
 is maintained throughout the disc evolution, it is expected that
  the collimation angle  of the jet is as small as $\leq 10^{\circ}$.  
  Furthermore, in addition to models 1 and 2, we examined the intermediate case  
  with $\dot m = 10$ between them and found that the case also  shows 
 anisotropic outflows  same as model 2 but with a smaller collimation angle 
 than  that in model 2. From these results, we speculate that
   the collimation angle may be smaller if we consider other models with 
  much higher input-accretion rates.  This should be confirmed in future.
  Of course, axisymmetric MHD disc simulations have been used 
  successfully in the past to reproduce jets formation and collimation of the 
 jets since \citet{b26}, and the recent 3-D MHD simulations, which have 
 directed at the internal dynamics of the disc itself and the resulting 
 accretion flow, also show an unbound outflowing jet along the Z axis confined 
 by magnetic pressure. However, as far as the very luminous X-ray source SS 433
  is concerned, the jet acceleration is  considered to be mainly in thermal 
  origin but not nonthermal one.  Therefore it is very 
  advantageous for us to be able to reproduce a small collimation angle of 
 the jet in terms of the thermal model considered here. 
  
  The mass-outflow rate in  model 1 is as large as  $\sim 10^{19}$ -- 
   $10^{23}$ g s$^{-1}$ but is variable and intermittent with times, 
   while in model 2 it is as large as $\sim 3 \times 10^{19}$ g s$^{-1}$,
    and it is stable and comparable
    to the mass-outflow rate estimated in SS 433.

\section{Concluding Remarks}
 Black-hole accretion discs and jets with 3 and 15 $\dot M_{\rm E}$
 have been examined by time-dependent two-dimensional radiation hydrodynamical 
 calculations over a far distant region and a long time-scale. 
 Starting with the initial disc based on the Shakura-Sunyaev
 disc model, we examined  the time evolutions of the disc and the jet.  
 As the result, the inner disc becomes geometrically thick, optically 
 thin, and convectively unstable.
The dominant radiation pressure force in the inner region of the disc
accelerates the gas flow vertically to the disc plane, and the high-velocity
 jets with 0.2 -- 0.4$c$ are formed along the rotational axis.
In model 1 with the lower accretion rate, the initially anisotropic 
high-velocity jet becomes gradually isotropic flow in the distant region of
 $\sim 10^{11}$ cm. The mass outflow occurs through the entire outer 
 boundary and the mass-outflow rate is large as $\sim 10^{19}$ -- $10^{23}$ 
 g s$^{-1}$, but it is variable and intermittent with times; that is, 
 the outflow  switches occasionally to inflow.
 At the initial phase the  luminosity fluctuates around $\sim 10^{40}$ 
erg s$^{-1}$ by a factor 2 and then it flares up to $4 \times 10^{40}$ 
  erg s$^{-1}$ and 
 again to  $ \sim 10^{42}$ erg s$^{-1}$ after a time of $2.5 
 \times 10^{5} r_{\rm g}/c$.  After then the luminosity behaves like 
  a periodic variation with a period of $\sim 2.5\times 
 10^{5}r_{\rm g}/c $ and an amplitude of factor ten.  
 
 On the other hand, the jet in model 2 with the higher accretion rate maintains its initial 
anisotropic shape even after the jets go far away, and the mass outflow occurs
 only through the polar direction.  The disc luminosity and the mass-outflow 
 rate attain to their steady values 
 of  $ 1.3 \times 10^{40}$ erg s$^{-1}$ and $3 \times 10^{19}$ g s$^{-1}$, 
 respectively. 
 These different behaviors of the unstable disc luminosity in model 1 and 
 the stable disc luminosity in model 2 are interpreted in terms of the
 thermal instability of the slim accretion disc model.
 The super-Eddington model with 15 $\dot M_{\rm E}$ 
 is promising to explain  a small collimation degree of 
 the jet and a large mass-outflow rate observed in the X-ray source SS 433.

 It shoud be noticed that these conclusions are dependent on the viscosity 
 parameter $\alpha$ because we examined only the low-viscosity flows with 
 $\alpha=10^{-3}$.  In the previous numerical simulations, we found that the 
 low-viscosity flow is more advantageous to form a well collimated powerful 
 jet than the high-viscosity flow with $\alpha=0.1$. 
 The characteristics of the accretion flows for 
 different $\alpha$ can qualitatively explained by the role of viscosity in the
 suppression of the convective instability in the accretion flows.
 Thick accretion discs of low viscosity ( $\alpha \leq 0.01$) are generally 
 convection-dominated flows \citep{b28,b10,b18,b24}.  The convective 
 instability results in the vortex motion of gas of different spatial scales.
 Convection cells stretch from the disc mid-plane to the disc surface in the
 inner region of the disc. The convective flow is subsonic in the mid-plane
 but becomes comparable to the sonic speed near to the disc surface.
 As a result, a part of the convective flow escapes from the disc as the disc 
 wind, dissipating energy of powerful accoustic waves produced by convection 
in the disc.  The outflowing material originates not only from the innermost
 disc near the transition region between the inner advection-dominated and 
 outer convection-dominated zones but also from the outer upper disc as
 the disc wind.  The disc wind is incorporated into the rarefied, hot, and 
 high-velocity region and is acclerated due to the strong radiation field
 in the rarefied region. However, in the case of a high viscosity, the strong
 accretion flow smoothes the convective instability and is globally stable to
 the instability.  As the result, the powerful jet as is found in the 
 low-viscosity case is not reproduced.  At present, we have no clear 
 conclusion as to which viscosity case to be realized in actual black hole
 accretion flows. 
 Future 3-D MHD 
 simulations of the black hole accretion flows may give a decisive answer to
 the viscosity problem.

 \section*{Acknowledgments}

 The authors would like to thank the referee for many useful comments.

\label{lastpage}

\end{document}